\documentclass[useAMS,usenatbib]{mn2e}
\usepackage[english]{babel}
\usepackage{amssymb}
\usepackage{natbib}
\usepackage{times}
\usepackage{graphics,epsfig}
\usepackage{graphicx}
\usepackage{amsmath}
\usepackage{supertabular}
\usepackage{mathrsfs,amsmath}
 
\usepackage{color}

\newcommand{\kmskpc}{\ensuremath{\rm km \thinspace s^{-1} \thinspace kpc^{-1}}}
\newcommand{\kmsarcsec}{\ensuremath{\rm km \thinspace s^{-1} \thinspace arcsec^{-1}}}

\newcommand{\kms}{km~s$^{-1}$}

\newcommand{\ha}{\ensuremath{{\rm H}\alpha}\ }

\newcommand{\ms}{\ensuremath{\rm M_{\odot}}}

\newcommand{\hh}{\ensuremath{\rm H_{2}}}

\newcommand{\mspcc}{\ensuremath{\rm M_{\odot} pc^{-3} }}

\newcommand{\HI}{H{\sc i}}

\begin{document}

\title[Slow \HI~bar in dwarf irregular galaxy DDO 168]{Detection of a slow \HI~bar in the dwarf irregular galaxy DDO 168}
\author[Patra \& Jog]{	Narendra Nath Patra$^{1}$ \thanks {E-mail: narendra@rri.res.in} and Chanda J. Jog$^{2}$ \\
	$^{1}$ Raman Research Institute, C. V. Raman Avenue, Sadashivanagar, Bengaluru 560080, India\\	
	$^{2}$ Department of Physics, Indian Institute of Science, Bengaluru 560012, India
}
\date {}
\maketitle

\begin{abstract}

We examine the \HI~total intensity maps of the VLA LITTLE-THINGS galaxies and identify an \HI~bar in the dwarf irregular galaxy DDO 168 which has a dense and compact dark matter halo that dominates at all radii. This is only the third galaxy found to host an \HI~bar. Using the \HI~kinematic data, we apply the Tremaine-Weinberg method to estimate the pattern speed of the bar. The \HI~bar is found to have an average pattern speed of $23.3 \pm 5.9$ \kmskpc. Interestingly, for the first time, we find that the observed pattern speeds of the bar in the two kinematic halves are different. We identify the origin of this difference to be the kinematic asymmetry. This observed offset in the pattern speed serves to put a stringent constraint on the lifetime of the bar set by the winding timescale. The lifetime of the bar is found to be $\sim 5.3 \times 10^8$ years, which is $\sim 2$ times the local dynamical time scale of the disc. We also find the \HI~bar in DDO 168 to be a weak bar with a strength of $\sim 0.2$. If \HI~bar being weak can be easily disturbed, this could possibly explain why it is extremely rare to observe \HI~bars in galaxies. We estimate the bar radius to be $\sim 1$ kpc and the dimensionless ratio, $R_L/R_b$ to be $\geq$ 2.1 indicating a `slow' bar in DDO 168. Our results confirm the proposition that the dynamical friction with the halo slows down a rotating bar in a galaxy dominated by dark matter halo from inner radii.

\end{abstract}


\section{Introduction}
\label{intro}

Bars are ubiquitous in spiral galaxies and nearly 75\% of galaxies in the local universe host bars in them \citep{eskridge00,masters11}. Further, the bar fraction is higher in low mass, late-type bluer galaxies compared to in high-mass, early-type redder galaxies \citep{nair10}. Since bars are strong, non-axisymmetric features, they are known to be major drivers of internal secular evolution of spiral galaxies \citep[e.g.,][]{sellwood93,sellwood14}. Bars can cause an outward transfer of angular momentum \citep[e.g.,][]{binney87} and inflow of gas resulting in an enhanced nuclear star formation \citep{aguerri99} or fuelling the central AGN \citep{shlosman89,friedli93}. Dynamically, it could also cause the vertical heating of stellar discs as well \citep{saha10}. Thus, bars play a crucial role in galaxy evolution, and hence, it is essential to understand their dynamics and physical properties.

Bars are instabilities, seen in the inner region of solid body rotation of a galaxy and their luminosity distribution can be well described by an $\rm m=2$  Fourier component. In N-body simulations as well as in the observed kinematics, a bar is seen to rotate like a solid body with a single pattern speed, $\Omega_p$. Despite extensive studies of bars, the physical mechanism responsible for their origin and their specific properties such as strength, length, and pattern speed are not well-understood \citep[e.g.,][]{binney87,sellwood14}. It is, however, well-known from numerical studies that a disc-alone simulation is violently unstable to the growth of a bar in it, while a dark matter halo tends to stabilize a disc against the growth of a bar.

A bar rotates with a constant pattern speed that is less than the angular speed of rotation of the disc. In the rotating frame of a strong perturbation like a bar, the Lagrangian point defines a location where a particle is stationary, which is equivalent to a co-rotation point in a general rotating frame. Theoretical and numerical studies reveal that a steady-state bar cannot extend beyond the Lagrangian point \citep{debattista00,sellwood06,athanassoula80,contopoulos80}.

Traditionally, a dimensionless quantity $\rm {\mathcal R} = R_L/R_b$, which is the ratio of the Lagrange point radius to the bar radius is found to have a value of $\sim 1 - 1.4$  in most early-type or HSB galaxies \citep{aguerri15,peters19}. Such bars are called ``fast'' bars \citep{debattista98}. However, if the dark matter halo of the host galaxy is dense, then the wake created in the halo by the rotating bar can lead to a dampening of the bar pattern speed, which effectively increases $R_L$ and hence, such bars could have $\rm {\mathcal R} \gg 1$, these are called slow bars \citep{debattista98,tremaine99}. \citet{debattista98} have shown that most HSB galaxies have fast bars and proposed that late-type or LSB galaxies where the dark matter halo is dense are likely to host slow bars. However, recent studies suggest that the dynamical braking between the bar and the dark matter halo could be effective in general for all galaxies, which can produce slow bars in HSBs as well \citep[see, e.g.,][]{font17}. 

Direct measurement of the bar pattern speed is difficult, particularly in faint, late-type galaxies. Many early studies adopted model-based pattern speed estimation from the photometric only data adopting $\Omega_p$ as a free parameter \citep{sanders80,hunter88,england90,garcia-burillo93,sempere95,lindblad96a,lindblad96b,laine99,aguerri01,weiner01,perez04,rautiainen08,treuthardt08}. Other studies used special morphological features (e.g., position of galaxy rings, sign-reversal of streaming motions across the corotation point etc.) extensively to estimate $\Omega_p$ in galaxies \citep{buta86,canzian93,buta95,vega97,munoz-tunon04,perez12}. However, these methods compare the morphological outcomes depending on the pattern speed and are model dependent. A model-independent direct measurement of pattern speed in galaxies using the photometry and kinematics is proposed by \citet[][hereafter TW]{tremaine84}. Several previous studies extensively applied the TW method to a large number of survey galaxies to estimate the pattern speed of bars in galaxies (e.g., SDSS-IV MaNGA survey \citep{bundy15} IFU data \citep{guo19}, the CALIFA survey \citep{sanchez12} data \citep{aguerri15} etc.). 

Recently, \citet{font11b} have proposed an innovative method (Font-Beckman method) to estimate the resonant radii in galaxies using the signature of the phase reversal of streaming motions, which is observed in the residual map of the velocity field. They used this method to estimate the corotation radii and hence the pattern speed in a large number of galaxies using \ha~data obtained with Fabry-Perot interferometry \citep{font17}. Further, they find this method to be consistent with the results from simulation \citep{font14c} as well as with the results produced by other methods, e.g., the TW method \citep{beckman18}. However, the Font-Beckman method requires high spatial and spectral resolution, which often is challenging to achieve in \HI~observations. Thus the TW method remains a promising method to estimate the pattern speeds where a high spatial and spectral resolution is not available. 


In this paper, we consider the dwarf irregular galaxy DDO 168 for which dark matter dominates at all radii \citep[see][for more details]{ghosh18}. We identify an \HI~bar in this galaxy and measure its pattern speed by applying the TW method to the \HI~kinematic data, from the VLA LITTLE-THINGS survey \citep{hunter12}. This study is of particular interest as the gaseous bars are rare, and it is only the third galaxy for which the pattern speed of the \HI~bar is measured in this paper. Further, the evaluated pattern speed in DDO 168 shows a slow bar, as would be expected from the dark matter dominance. This supports the idea proposed by \citet{debattista98} that a dense dark matter halo can effectively slow down the rotating bar through dynamical friction. We find other interesting features such as non-equal pattern speeds in the two halves of the bar, which we argue could arise due to the kinematic asymmetry observed in the galaxy. Furthermore, a simple dimensional argument allows us to put an upper limit to the age of the bar. This, in turn, could help in explaining the paucity of gaseous bars in galaxies.

\section{The Tremaine-Weinberg method}

Using the observed kinematic and photometric data, \citet{tremaine84} proposed a direct method to estimate the pattern speeds of bars in galaxies. This method does not count on specific modeling, rather explicitly depends only on observables. However, it relies upon the assumption that the tracer population tracing the kinematics/pattern, follows the continuity equation. However, though this assumption works fairly well for an old stellar population which traditionally being used to investigate bars in galaxies, the gas tracers (\HI~or \hh) particularly might deviate from this assumption due to the conversion between atomic and the molecular phases. Nonetheless, the fractional conversion between these phases is minimal. For example, in dwarf galaxies (whose ISM is dominated by \HI), no significant molecular gas is observed till date \citep{israel95,taylor98b,schruba12}. 
Further, the star formation efficiency in these galaxies is found to be poor in contrast to the large spirals, even where the conversion of gas into stars is $\lesssim$ a few percents per $10^8$ years \citep{kennicutt98b}. In this sense, the gas tracers are expected to satisfy the continuity condition reasonably well and thus give a reliable estimate of pattern speeds. Many previous studies used the gas as a tracer to estimate the pattern speeds in galaxies using the TW method \citep{rand04,banerjee13,speights12}. For DDO 168, the molecular gas content is negligible and the star formation rate is low, $\sim 6.3 \times 10^{-3} \ M_\odot yr^{-1}$ \citep{hunter04} making it a plausible target for pattern speed estimation of its \HI~bar.

Following \citet{tremaine84}, for a tracer population following the continuity equation, a well-defined pattern speed, $\Omega_p$, can be given as 

\begin{equation}
\Omega_p \sin(i) = \frac{\int_{-\infty}^{+\infty} h(y) \int_{-\infty}^{+\infty} \Sigma(x,y)v_r(x,y) dx dy}{\int_{-\infty}^{+\infty} h(y) \int_{-\infty}^{+\infty} \Sigma(x,y) dx dy}
\label{eq1}
\end{equation}

\noindent where (x,y) represents Cartesian coordinates of the tracer in the sky plane. The origin of this coordinate system is coincident with the center of the galaxy, and the x-axis is aligned to the line-of-nodes, i.e., the major axis of the galaxy. The $\Sigma(x,y)$ is the surface density of the tracer population (\HI~in this case) and $v_r(x,y)$ is the observed radial velocity (after subtracting the systematic velocity) at (x,y). $i$ represents the inclination of the galaxy disc and $h(y)$ describes an appropriate weighting function which can arbitrarily be chosen to decide the pattern speed calculation at a particular $y$. A Conventional form of $h(y) = \delta(y-y_0)$ represents slits or pseudo-slits parallel to the line-of-nodes \citep[see][for more details]{tremaine84}.

Eq.\ref{eq1} can be rewritten as 

\begin{equation}
\Omega_p \sin(i) = \frac{\langle V \rangle}{\langle X \rangle}
\label{eq2}
\end{equation}

\noindent where,
\begin{equation}
\langle V \rangle = \int_{-\infty}^{+\infty} h(y) \int_{-\infty}^{+\infty} \Sigma(x,y)v_r(x,y) dx dy
\label{eq3}
\end{equation}
\noindent and, 

\begin{equation}
\langle X \rangle = \int_{-\infty}^{+\infty} h(y) \int_{-\infty}^{+\infty} \Sigma(x,y) dx dy
\label{eq4}
\end{equation}

The above integrals are called TW integrals and could be evaluated along many $y$ values, i.e., slit/pseudo slit positions. If the bar has a meaningful, constant pattern speed, a $\langle V \rangle$ vs. $\langle X \rangle$ plot is expected to result in a straight line (due to constant $\Omega_p$). The slope of the straight line then can be used to estimate $\Omega_p$. It should be noted that if there is no clear pattern speed present or it acquires different values at different positions on the bar, $\langle V \rangle$ vs. $\langle X \rangle$ curve is expected to deviate from a straight line \citep[see, e.g.,][]{rand04}. In other words, the existence of a well-defined pattern speed can be asserted if the $\langle V \rangle$ vs. $\langle X \rangle$ relation follows a reasonable straight line (see \S~\ref{res} for more details).


\begin{figure*}
\begin{center}
\begin{tabular}{c}
\resizebox{0.99\textwidth}{!}{\includegraphics{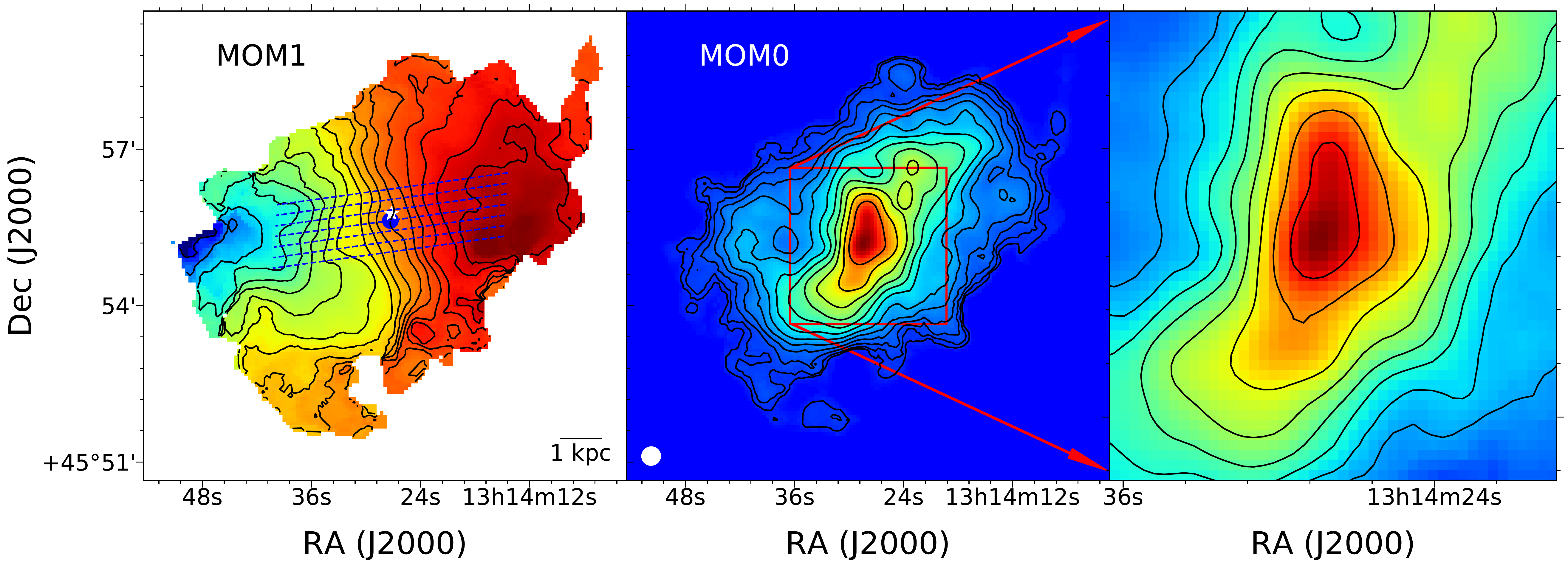}}
\end{tabular}
\end{center}
\caption{Left panel: The velocity field (MOM1) of DDO 168. The contours vary from 140 to 225 \kms~in step of 5 \kms. The blue filled circle at the center indicates the kinematic center of the galaxy \citep{oh15} whereas the white asterisk represents the photometric center \citep{hunter12}. However, the difference is less than a third of the observing beam as shown by the white circle at the bottom left corner of the middle panel. The blue dashed lines in the left panel indicate the representative slit positions used to calculate $\Omega_p$. Middle panel: \HI~column density map of DDO 168 as obtained from the LITTLE-THINGS survey. The contour levels are $(8., 11.2, 16., 22.4, ....) \times 10^{19} \ atoms \thinspace cm^{-2}$. The \HI~bar can be seen prominently at the center. For better clarity, we zoom the central region of the galaxy (marked by the red box) in the right panel. See text for more details.}
\label{mom}
\end{figure*}

DDO 168 is a gas-rich dwarf irregular galaxy with a highly-extended \HI~disc. In table~\ref{tab:prop} we present some basic properties of this galaxy. As can be seen, neutral gas in this galaxy dominates the baryonic mass budget with a gas mass $\sim 5$ times higher than the stellar mass \citep{johnson15}. Interestingly, DDO 168 is also an extremely dark-matter-dominated galaxy with a $M_{DM}/M_{baryon} \sim 10$ \citep{johnson15} and has a compact dark matter core \citep{ghosh18}. This indicates, the dark matter dominates the dynamical mass right from the center of the galaxy. This, in turn, makes this galaxy an ideal site where one can test the proposition that a dense dark matter halo slows down the rotating bar through dynamical friction. 

\begin{table}
\caption{Basic properties of DDO 168}
\begin{center}
\begin{tabular}{lcr}

\hline
Parameters & Value & Ref \\
\hline
$^a$RA (J2000) &  13h14m27.2 & \\
$^a$DEC (J2000) &  +45d55m46 & \\
$^b$RA (J2000) & 13h14m27.3s & 1\\
$^b$DEC (J2000) & +45d55m37.3 & 1\\
Distance (Mpc) & 4.3 & 2 \\
$M_V$ (mag) & -15.7 & 6 \\
$v_{sys}$ (\kms) & 192.6$\pm$1.2 & 1\\
Inclination, $i (^o)$ & 54.5 & 5 \\
Stellar mass, $\log M_* (M_\odot)$ & 7.73 & 3 \\
Total gas mass, $M_{gas}$ ($\times 10^7$ \ms) & 25.94 & 1\\ 
Dynamical mass, $\log M_{dyn} (M_\odot)$ & 9.52 & 1 \\
B-band Holmberg radius, $R_H$ (arcmin) & 2.32 & 4 \\ 
Photometric PA of the \HI~disk ($^o$) & -24.5 & 5 \\ 
Kinematic PA of the \HI~disk ($^o$)& 275.5 & 5 \\ 

\hline

\end{tabular}
\end{center}
\label{tab:prop}

$^a$Photometric center, $^b$Kinematic center, $^1$\citet{oh15}, $^2$\citet{karachentsev03b}, $^3$\citet{johnson15}, $^4$\citet{hunter06}, $^5$\citet{hunter12}, $^6$\citet{hunter99}
\end{table}

\section{Results and discussions}
\label{res}

\begin{figure*}
\begin{center}
\begin{tabular}{c}
\resizebox{0.99\textwidth}{!}{\includegraphics{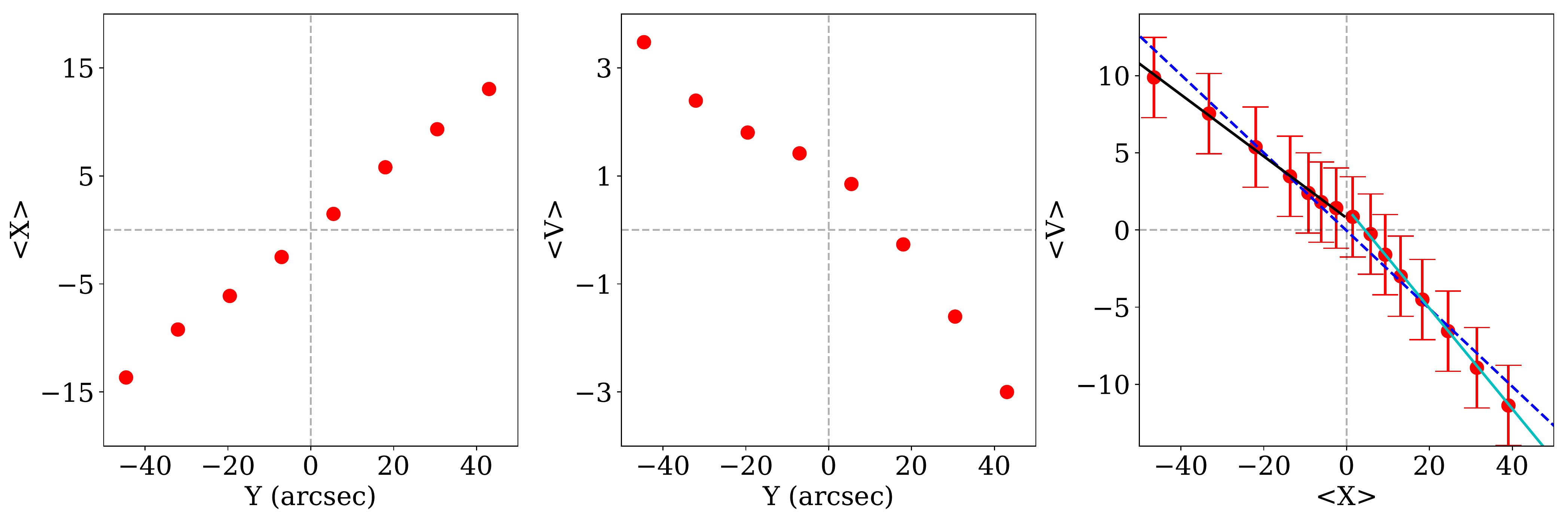}}
\end{tabular}
\end{center}
\caption{Left panel: The TW integral, $\langle X \rangle$ as a function of different slit position in arcsec. The representative slits (blue dashed lines) used to calculate the TW integrals, the slits are separated by $\sim 1/3$ of the observing beam. Middle panel: The TW integral $\langle V \rangle$ as a function of slit position in arcsec. Right panel: $\langle V \rangle$ vs. $\langle X \rangle$ plot. The blue dashed line represents a linear fit to the data points. Assuming an inclination of 54.5$^o$ \citep{hunter12}, the estimated pattern speed is $\Omega_p = 23.3 \pm 5.9$ \kmskpc. However, it can be seen that there is a change in the slope in the $\langle V \rangle$ vs. $\langle X \rangle$ plot at $\langle X \rangle \sim 0$ which indicates different pattern speeds in the two halves. Individual points in both the halves are fitted with separate straight lines as shown by the solid black and cyan lines.  See text for more details.}
\label{pspeed}
\end{figure*}
\subsection{Identification of the \HI~bar }
As mentioned in \S~2, the line-of-sight velocity, i.e., $v_r(x,y)$ and the surface density of \HI, i.e., $\Sigma(x,y)$ are the primary inputs to Eq.~\ref{eq3} and~\ref{eq4} for computing the pattern speed. In Fig.~\ref{mom} left panel we show the velocity field of DDO 168 derived using the \HI~spectral cube from the LITTLE-THINGS survey. A clear velocity gradient can be seen in color scale across the galaxy indicating the existence of an ordered rotation. Moreover, the velocity field displays the signature of a typical rigid rotator as observed in other dwarf galaxies as well \citep[see, e.g.,][]{oh15}. The middle panel of the figure shows the \HI~column density map of DDO 168. An \HI~bar can clearly be seen at the center of the galaxy which is almost vertical to the map. For further emphasis, we zoom in the central region in the next panel to the right to show the \HI~bar more clearly. However, it should be noted that due to its gas nature, the orbits of the \HI, forming the bar are more susceptible to external perturbations than stars. A signature of the same is visible in the two rightmost panels of Fig.~\ref{mom}. The observed \HI~bar is not completely symmetric and 
rectangular. Further, from the middle panel, it can be seen that two tentative spiral arms are originating near the bar ends which is very similar to what is observed in large barred spiral galaxies.

\subsection{Determination of the bar pattern speed}
For the estimation of the pattern speed of the bar in DDO 168, the kinematic and photometric parameters are essential. However, it is found that some of these parameters acquire different values depending on 
whether a kinematic or a photometric analysis is carried out. For example, the photometric and kinematic centers for DDO 168 are not coincident. In the left panel of Fig.~\ref{mom} the blue circle at the center represents the kinematic center of DDO 168 \citep{oh15} whereas the white asterisk indicates the photometric center. However, the difference is less than $1/3$ of the observing beam ($\sim 21^{\prime \prime}$) and hence it does not have a considerable impact on the pattern speed calculation. There is also a significant ($\sim 60^o$) difference between the kinematic and the morphological position angles (PAs) as well. The position angle plays a vital role in defining the lines-of-nodes and hence greatly influences the calculated pattern speed 
\citep{debattista03}. As position angle decides the lines-of-nodes, it must separate the kinematic halves and hence, we use the kinematic PA as the line-of-nodes for DDO 168 \citep[see][for a similar case]{banerjee13}. In fact, for DDO 168, adoption of the photometric position angle (-$24.5^o$; \citet{hunter12}) as the lines-of-nodes do not produce a meaningful pattern speed of the \HI~bar, i.e., a $\langle V \rangle$ vs. $\langle X \rangle$ plot does not exhibit a reasonable straight line. In the left panel of Fig.~\ref{mom} we show the representative slits (blue dashed lines) we use to calculate the TW integrals. For DDO 168, we use slits separated by $\sim 1/3$ of the observing beam. The other physical parameters of DDO 168 used to estimate the pattern speed of its \HI~bar are listed in Tab.~\ref{tab:prop}.

Using the \HI~velocity and the intensity maps, we evaluate Eq.~\ref{eq3} and~\ref{eq4} along the slits as shown in Fig.~\ref{mom}. In Fig.~\ref{pspeed} left and middle panels, we plot the corresponding TW integrals for DDO 168. Consequently, $\langle V \rangle$ vs. $\langle X \rangle$ is plotted in the rightmost panel. As mentioned earlier, the slope of the fitted straight line to these data points would provide us the pattern speed of the \HI~bar. Hence, we fit the data points with a straight line (blue dashed line in the right panel) and find a slope, $m=-0.25 \pm 0.01$ \kmsarcsec~with an intercept, $c=-0.05 \pm 0.16$ \kms. With these values, we obtain the pattern speed of the bar to be, $\Omega_p = 23.3 \pm 5.9$ \kmskpc~adopting a distance of 4.3 Mpc for DDO 168. The dominant contribution to the uncertainty in the pattern speed comes from the error in the assumed position angle. Adopting a similar approach as used by earlier studies, we use a $\pm 5^o$ variation in the PA to determine the error bar on $\Omega_p$.

In the right panel of Fig.~\ref{pspeed}, we note that the $\langle V \rangle$ vs. $\langle X \rangle$ plot has a break at $\langle X \rangle \sim 0$. Consequently, the slopes of the $\langle V \rangle$ vs. $\langle X \rangle$ curves in the upper half and the lower half are different. This indicates two different observed pattern speeds in the two halves. We fit two straight lines (solid black and cyan lines) to the points in the two halves to calculate the pattern speeds. A pattern speed, $\Omega_{p+} = 18.4 \pm 5.7$ \kmskpc~is found for the upper half whereas, an $\Omega_{p-} = 30.0 \pm 7.2$ \kmskpc~is found for the lower half. As can be seen, the difference in the pattern speeds is significant and highly unlikely to be accounted for by measurement uncertainties. It should be mentioned here that, for DDO 168,  $\langle X \rangle = 0$ refers to the central slit (line-of-node) dividing the galaxy into two kinematic halves. We also emphasize that the TW integrals do follow a near straight line, indicative of  having a well-defined pattern speed in each half.

\subsection{Origin of asymmetry of bar pattern speed}
\begin{figure*}
\begin{center}
\begin{tabular}{c}
\resizebox{0.99\textwidth}{!}{\includegraphics{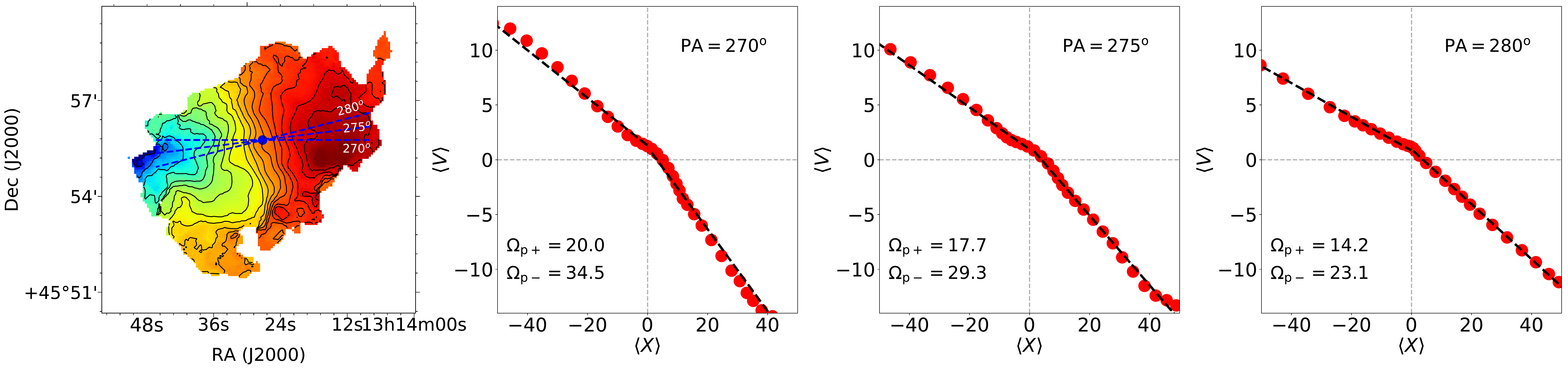}}
\end{tabular}
\end{center}
\caption{The pattern speeds in the two halves of DDO 168 are evaluated for different assumed position angles. Left panel: The assumed position angles are shown by the dashed blue lines on top of the velocity field of DDO 168. Different lines demarcate different lines-of-nodes corresponding to different position angles as quoted in the figure. The contours are the same as shown in the left panel of Fig.~\ref{mom}. The three panels on the right show the $\langle V \rangle$ vs. $\langle X \rangle$ plots for the three different assumed position angles as quoted on the top right of the individual panels. The red filled circles represent the measured points whereas the black dashed lines in each panel represent a linear fit to the data. The data were separately fit with straight lines in both the upper ($\langle X \rangle < 0$) and the lower halves ($\langle X \rangle > 0$). The corresponding pattern speeds ($\Omega_{p+}$ and $\Omega_{p-}$) are quoted in the bottom left corners of the respective panels in the units of \kmskpc. As can be seen, a change in the PA cannot eliminate the difference in the pattern speeds in the two different halves.} 
\label{pavar}
\end{figure*}

Next, we investigate the probable causes which could lead to two different pattern speeds at the two separate halves of the bar. We first examine the effect of the assumed position angle. As discussed earlier, for DDO 168, there is a significant difference between the kinematic and photometric position angles which might be a possible source of the observed pattern speed offset. To test the same, we evaluate the TW integrals and hence the pattern speeds for three different assumed position angles, i.e., $270^o$, $275^o$ and $280^o$ two of which are roughly $5^o$ apart from the observed kinematic position angle of $275.5^o$. In Fig.~\ref{pavar} left panel we show these three position angles by the blue dashed lines. In the rest of the panels, we show the corresponding $\langle V \rangle$ vs. $\langle X \rangle$ plots. We again fit the $\langle V \rangle$ vs. $\langle X \rangle$ points with straight lines in each half to estimate the pattern speeds as quoted in the bottom left of the respective panels. As can be seen from the figure, a change in the position angle does not eliminate the difference in the pattern speeds. To check the consistency, we also perform this exercise for a range of position angles starting from $250^o$ to $300^o$ with a step of $2^o$ and find no suitable position angle for which the pattern speeds in two different halves match with each other unquestionably. This exercise demonstrates that an error in the assumed position angle cannot possibly cause two different pattern speeds at two different halves of the gaseous bar in DDO 168. 

\begin{figure*}
\begin{center}
\begin{tabular}{c}
\resizebox{0.99\textwidth}{!}{\includegraphics{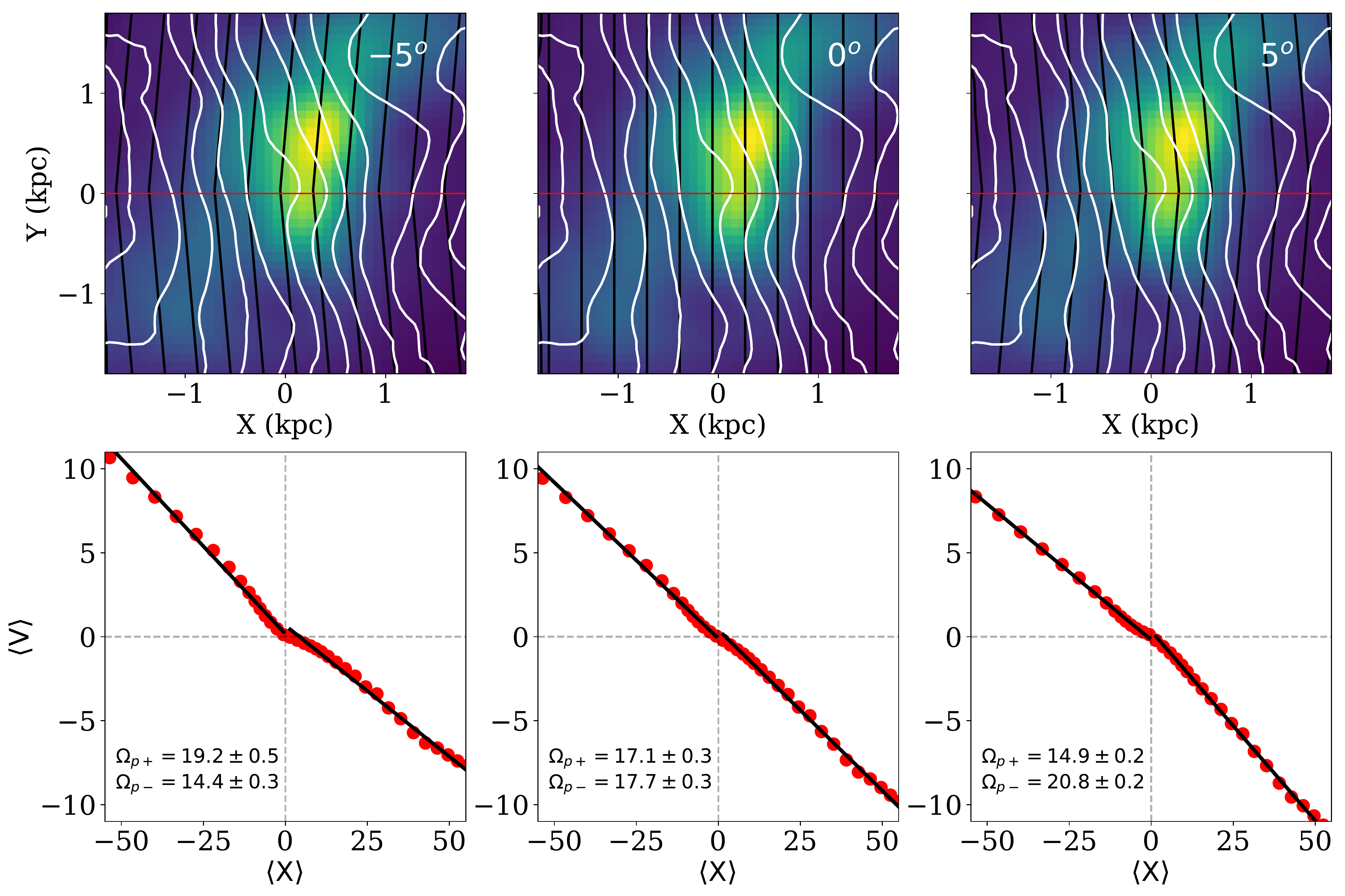}}
\end{tabular}
\end{center}
\caption{Top panels: The color scale represents the \HI~intensity distribution in the central bar region whereas the white contours show the observed velocity field. The black contours represent the simulated velocity field with an additional asymmetry introduced. The rigid body rotation velocity fields in the two halves have an offset as quoted by the angles at the top right of the respective panels. The solid red line at the center represents the line-of-node corresponding to a position angle of 275.5$^o$. Bottom panels: The estimated pattern speeds for the simulated velocity fields. The solid red circles describe the evaluated TW integral values whereas the solid black lines represent a linear fit to the data. Corresponding pattern speeds in both the halves ($\Omega_{p+}$ and $\Omega_{p-}$) are quoted at the bottom left corners of the individual panels. See text for more details.} 
\label{angvar}
\end{figure*}

We further explore the observed non-symmetric bar intensity distribution as a possible cause for the difference in the pattern speeds. To check that, we artificially generate a total \HI~intensity map with a symmetric, rectangular bar at the center. We use this map as $\Sigma (x,y)$ and evaluate the TW integrals to calculate the pattern speed. It should be noted that the observed velocity field is used so that the effect of the bar intensity only be examined. We find that a perfectly axisymmetric rectangular bar cannot eliminate the observed difference in the pattern speeds. This illustrates that a non-symmetric (or partially disturbed) nature of the \HI~bar cannot be attributed as a possible reason for the observed pattern speed difference in DDO 168. 

Next, we investigate the velocity field, i.e., $v_r(x,y)$ which is one of the critical input to Eq.~\ref{eq3} to evaluate the TW integrals. We artificially simulate the velocity field of DDO 168 such that it retains the total observed velocity gradient across the galaxy but with an angular offset between the two kinematic halves. This effectively produces a kinematic asymmetry. In other words, the rotation curves in the two halves produced by this simulated velocity field will have a slight offset. This kind of offset is not uncommon in observation. In fact, a significant fraction of galaxies are found to have this kind of kinematic asymmetry \citep{oh15}. In addition, a large fraction of galaxies is known to be lopsided (or with m=1 asymmetry) in spatial and kinematical properties \citep{jog09}. In Fig.~\ref{angvar} top panels we show this velocity field by black contours. Different panels represent a varying degree of offset. For example, in the middle panel, the offset is zero, and the iso-velocity contours are expected to run smoothly without any break (what is expected in a solid body rotation) across the kinematic axis (solid red line). The left and the right panels represent two different offsets where the angle between the iso-velocity field lines with respect to the zero offset are -$5^o$ and $5^o$ respectively. We use these velocity fields and the observed \HI~intensity map ($\Sigma (x,y)$) to evaluate the pattern speeds in two different halves using the TW method. The results are shown in the bottom panels of Fig.~\ref{angvar}. As can be seen in the left panel, when the velocity field lines are offset by -$5^o$, the pattern speeds in the two different halves are significantly different as quoted in the bottom left corner. A similar trend is also seen for a relative angle of $5^o$ too (right panel). However, when the angle is $\simeq 0$, i.e., there is no kinematic asymmetry, the pattern speeds in both the halves agree with each other very well as shown in the bottom middle panel of Fig.~\ref{angvar}. It should be mentioned here that the errors quoted on the pattern speed measurements in Fig.~\ref{angvar} only reflect the fitting errors and do not account for the uncertainty in the assumed position angle. We emphasize that, the relative value of the slopes of the $\langle V \rangle$ vs. $\langle X \rangle$ plots in both the halves changes sign across the panels from left to the right. For example, in the bottom left panel, the pattern speed of the upper half is higher than the lower half. Whereas in the rightmost panel, the pattern speed of the lower half is higher than the upper half. This suggests that there would be an asymmetry angle for which these pattern speeds will just cross each other, eliminating the break in the $\langle V \rangle$ vs. $\langle X \rangle$ plot. This exercise implies that the kinematic asymmetry is responsible for producing the observed offset in the pattern speed in DDO 168.

We note that to our knowledge such asymmetry for bar pattern speed has not been noted in the literature so far. Such asymmetry for the pattern speed of spiral arms on the two sides of a galaxy has been noted \citep{speights12} using TW method, and also a radial variation in the spiral pattern speed has been measured in some galaxies with a lower value at larger radii \citep{meidt09} but these could be explained due to swing amplification applied locally which could plausibly have different pattern speeds.

\subsection{Estimation of the lifetime of the bar}

Moreover, this difference in the pattern speed has significant physical implications in connection to the lifetime and stability of the gaseous bar. For a steady bar, it is expected to perform a solid body rotation with a single pattern speed. Various proposed mechanisms for the generation of bars such as global mode instability would be difficult to operate so as to give different pattern speeds on two sides. It is possible that the bar in DDO 168 started with a constant bar pattern speed, but some mechanism acting later on, such as a different torque on the two sides say due to gas infall or tidal interaction, could have caused the pattern speed values on the two sides to differ. Irrespective of what caused this difference in rotation speeds, one can ask how long such a system, namely the bar with different pattern speeds on two sides would survive? We estimate the winding time, i.e., the time for the bar to be smeared totally as 

\begin{equation}
t_{winding} = 2 \pi /(\Omega_{p+} - \Omega_{p-})
\end{equation}

\noindent with $\Omega_{p+} = 30.0$ \kmskpc~and $\Omega_{p-} = 18.4$ \kmskpc, we obtain $t_{winding} \sim 5.3 \times 10^8$ year. Thus this bar cannot last for a Hubble time, and this could be one of the reasons why the \HI~bars are so rare. On the other hand, note that the above winding time is equivalent to $\sim 2$ local dynamical time scale, $2 \pi/(\Omega_p) \sim 2.6 \times 10^8$ years. Hence, though it is disturbed, such a bar can last over a couple of dynamical timescales making it possible to be detected, albeit with a lower probability. Thus, there is a small but finite timescale over which a disturbed \HI~bar can be seen in galaxies such as DDO 168. 

\subsection{Determination of length of the \HI~bar}

Like pattern speed, the length of the \HI~bar is one of the critical physical parameters. However, unlike stellar bars, gaseous bars do not exhibit strongly ordered structures in general, and they are highly susceptible to external perturbations leading to disturbed morphology. Due to these reasons, it is hard to measure the length of a gaseous bar accurately. Adopting a similar approach presented in \citet{banerjee13}, we use three different methods, namely, i) {\it Visual inspection} ii) {\it Position angle variation} and iii) {\it Fourier decomposition} to estimate the length of the semi-major axis ($R_b$) of the \HI~bar in DDO 168.

\begin{figure}
\begin{center}
\begin{tabular}{c}
\resizebox{0.49\textwidth}{!}{\includegraphics{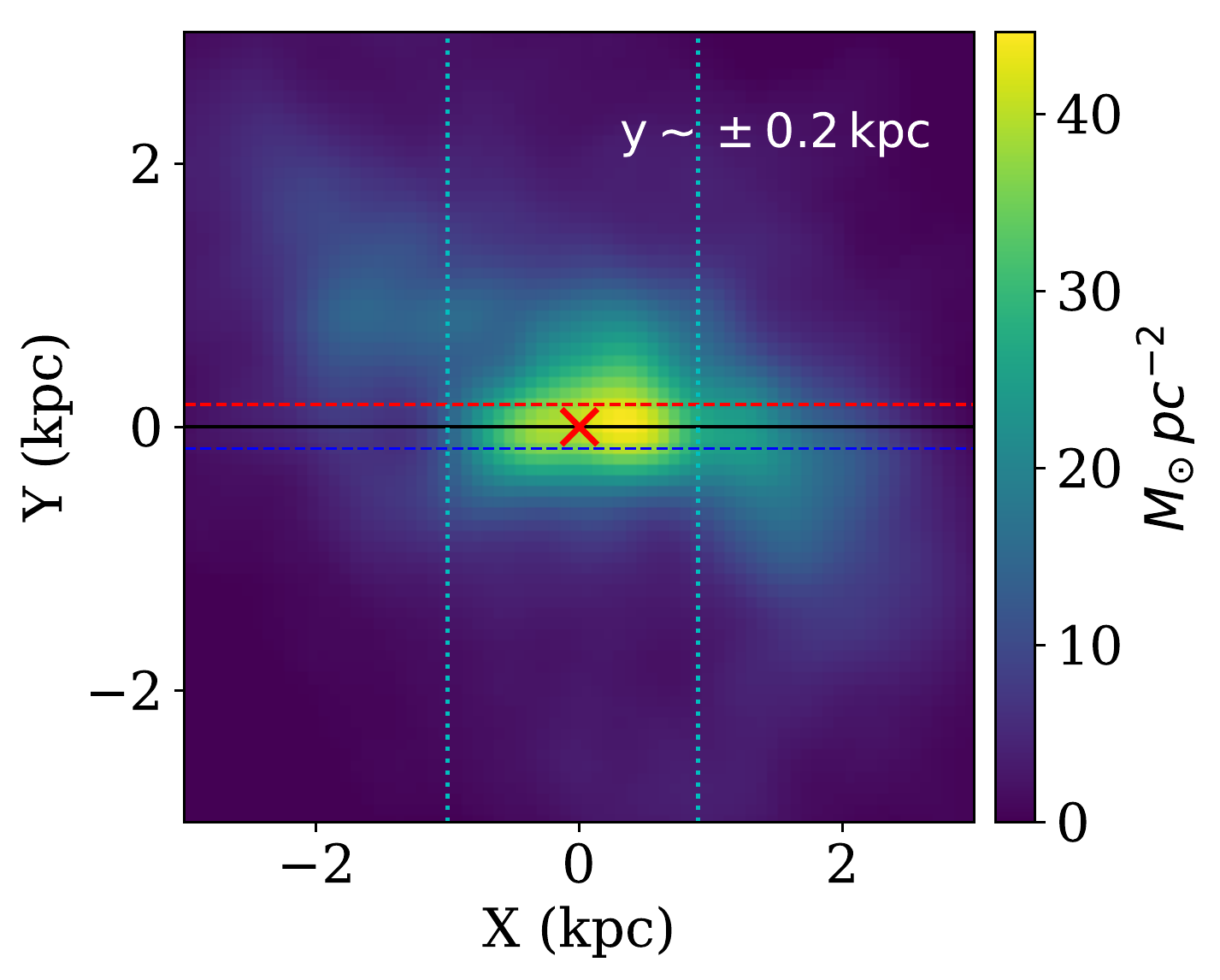}}\\
\resizebox{0.45\textwidth}{!}{\includegraphics{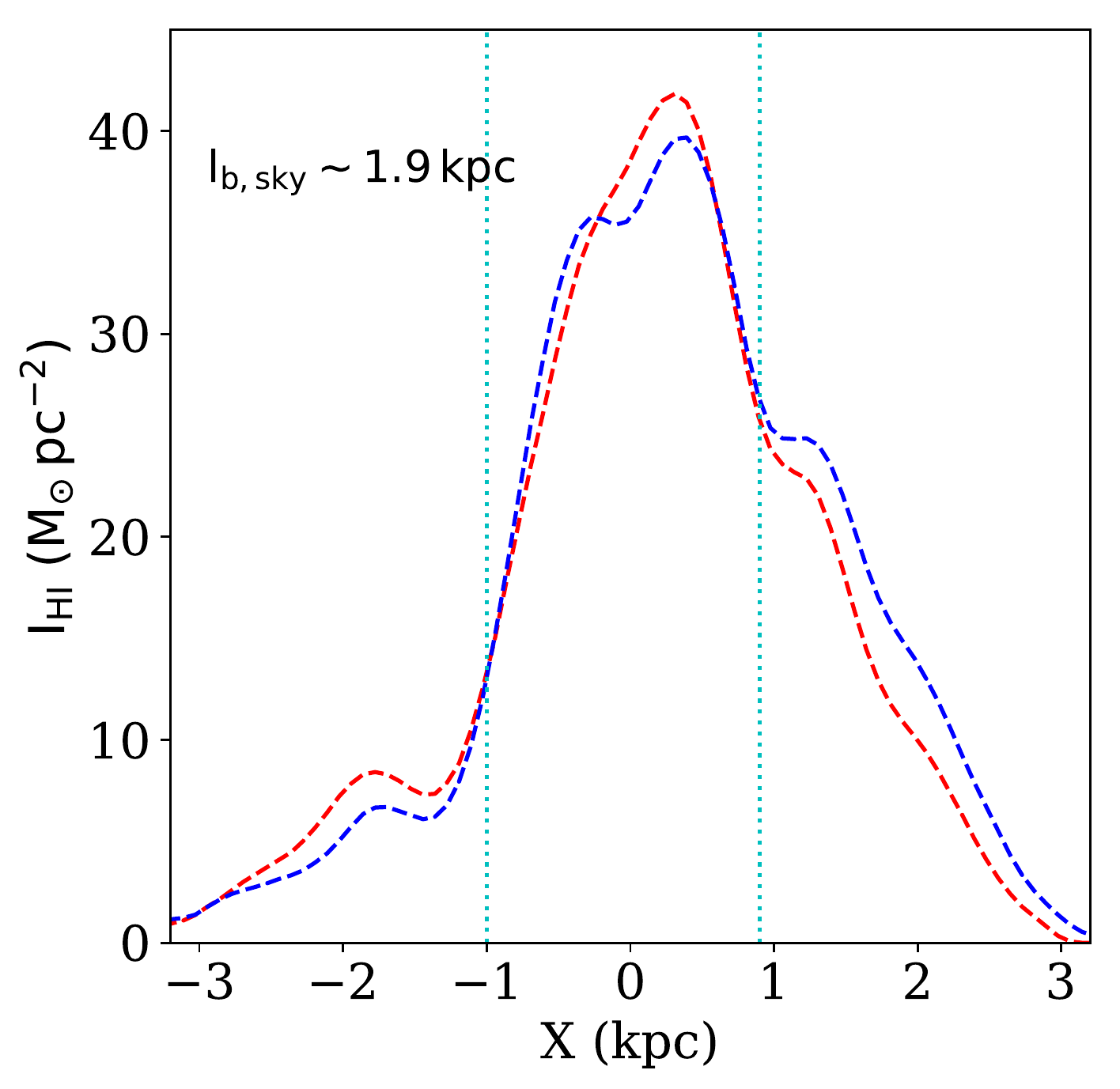}}
\end{tabular}
\end{center}
\caption{Estimation of the bar length through visual inspection. Top panel: The color scale shows the \HI~intensity distribution in the central region of DDO 168. The two cyan dotted lines demarcate the bar region. The red cross represents the morphological center of the bar whereas the solid black line represents the bar axis (PA = -$7^o$). The red and the blue dashed lines represent the trails (at $y \sim \pm 0.2$ kpc) along which the \HI~intensity profiles are extracted for visual inspection. These intensity profiles are shown in the bottom panel. The red and blue dashed curves are for the red and the blue lines in the top panel. The cyan lines in the bottom panel define the bar region (same as in the top panel) outside which the \HI~intensity drops significantly. Using this method we find the length of the bar in sky plane to be $l_{b,sky} \sim 1.9$ kpc.} 
\label{visual}
\end{figure}

\smallskip
\noindent{\it Visual inspection:} In this method, the bar region is visually inspected to assess the length of the \HI~bar in DDO 168. In Fig.~\ref{visual} top panel we show the \HI~intensity distribution of the central region in DDO 168. The \HI~bar can distinctly be seen in the total intensity map. To quantitatively estimate its extent, we extract intensity cuts along the major axis of the bar (blue and the red dashed lines). The intensity profiles thus generated are plotted in the bottom panel. It can be seen from the figure, that the intensities fall sharply as one moves out of the bar region. We then visually determine the boundary of the bar as demarcated by the cyan dotted lines in both the panels. We find the length of the semi-major axis of the bar in the sky-plane, $\rm l_{b,sky}\sim 1.9$ kpc. Assuming a bar position angle of $-7^o$ (see the {\it position angle} method), photometric position angle of $-24.5^o$ for the \HI~disc of DDO 168 \citep{hunter12} and an inclination of 54.5$^o$, we find the deprojected bar radius to be $\sim 1$ kpc.

\begin{figure}
\begin{center}
\begin{tabular}{c}
\resizebox{0.45\textwidth}{!}{\includegraphics{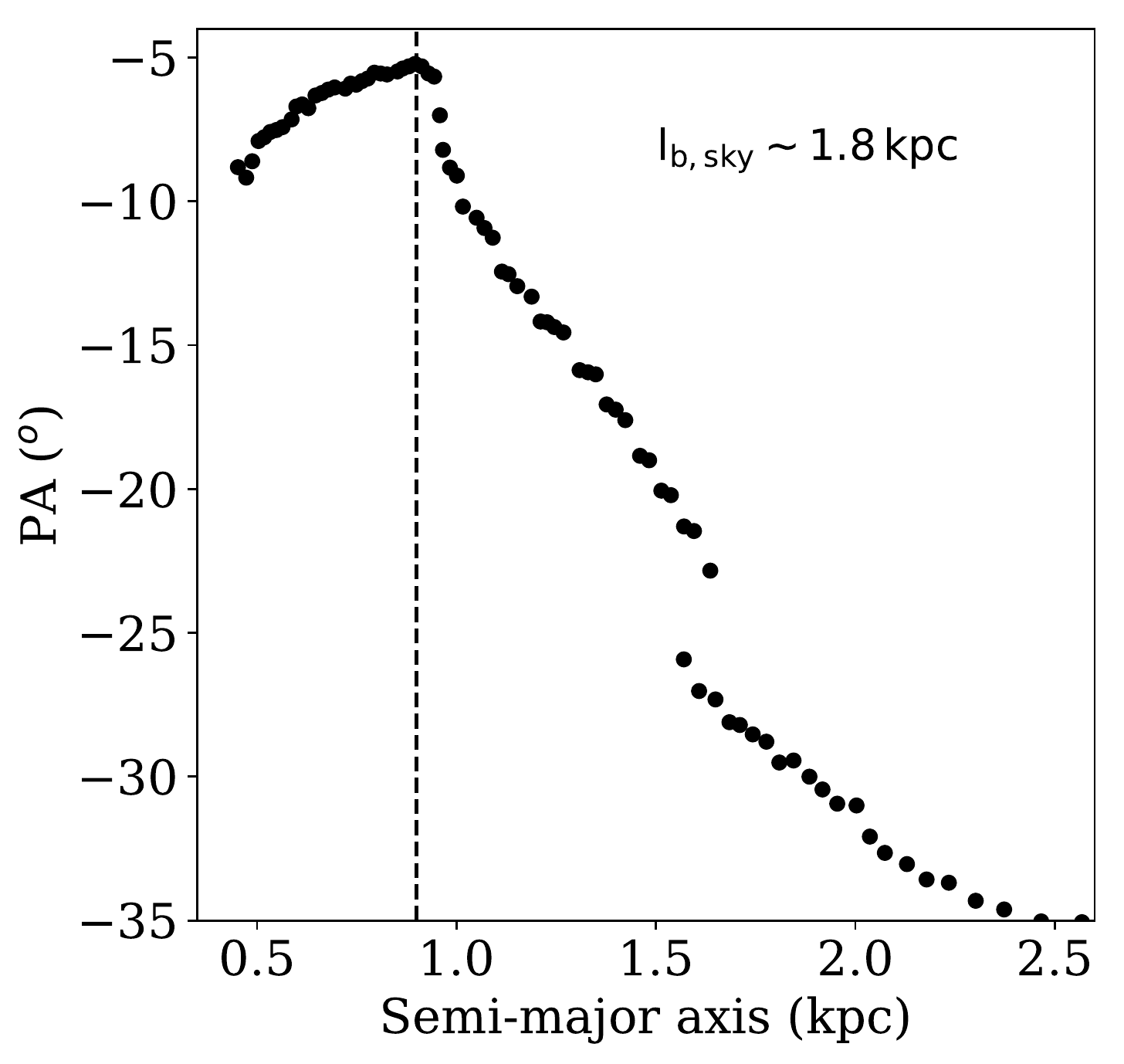}}
\end{tabular}
\end{center}
\caption{Estimation of bar length using the PA variation method. The PA of the fitted ellipses to the \HI~isophotes are plotted as a function of their semi-major axes length in kpc. The PA can be seen to vary rapidly beyond a semi-major axis length of $\sim 0.9$ kpc, shown by the black dashed line. Using this method, we estimate the length of the bar in the sky plane to be $\sim 1.8$ kpc.} 
\label{pa}
\end{figure}

\smallskip
\noindent{\it Position angle variation:} This is one of the traditional methods which is widely used to estimate the bar lengths in galaxies using stellar photometric data. In this method, the isophots are fitted with ellipses inside out from the central bar region. Then the position angles of these fitted ellipses are scrutinized as a function of their semi-major axis lengths. Because the bar is a well-defined regular structure, it is expected that within the bar region, all the isophots will acquire very similar position angles. However, outside the bar region, the isophots or the intensity distribution becomes irregular or patchy due to non-regular structures present in the galaxy. This results in a sudden change in the position angles of the fitted ellipses as soon as one moves out of the bar region. For DDO 168, we fit the \HI~iso-intensity contours from the central bar region out to a radius of $\sim 2.5$ kpc. In Fig.~\ref{pa} we plot these position angles as a function of semi-major axis length in kpc. As can be seen from the figure, at a radius of $\sim$ 0.9 kpc, there is a sudden change in the trend, and the PA starts deviating from a quasi-stable value of $\sim -7^o$. We adopt this as the bar radius (as shown by the black dashed line in Fig.~\ref{pa}) and estimate the deprojected bar radius to be $\sim 1$ kpc. It should be noted that the bar radius thus evaluated is consistent with what is obtained in the {\it visual inspection} method.

\begin{figure}
\begin{center}
\begin{tabular}{c}
\resizebox{0.45\textwidth}{!}{\includegraphics{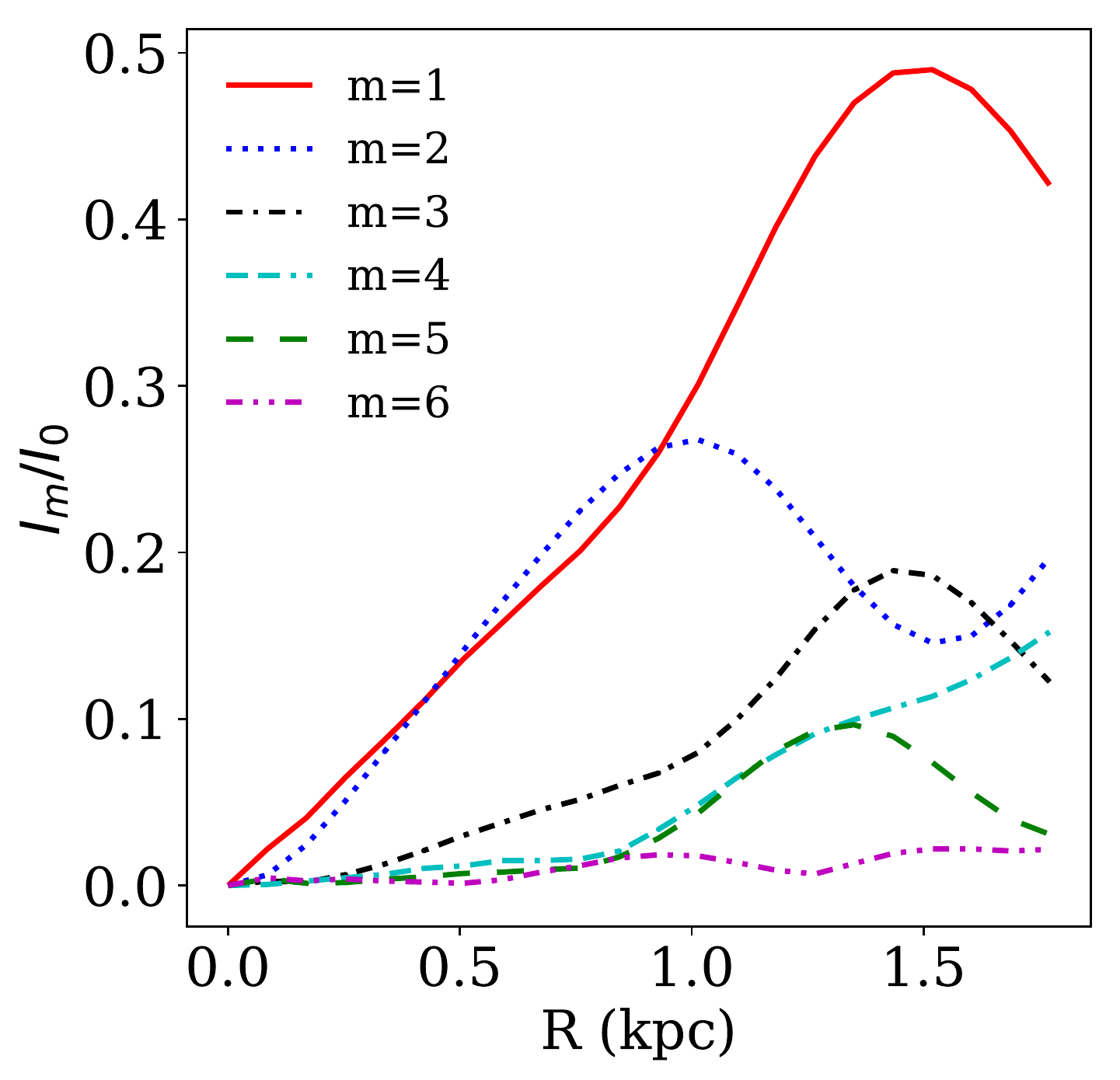}} \\
\resizebox{0.45\textwidth}{!}{\includegraphics{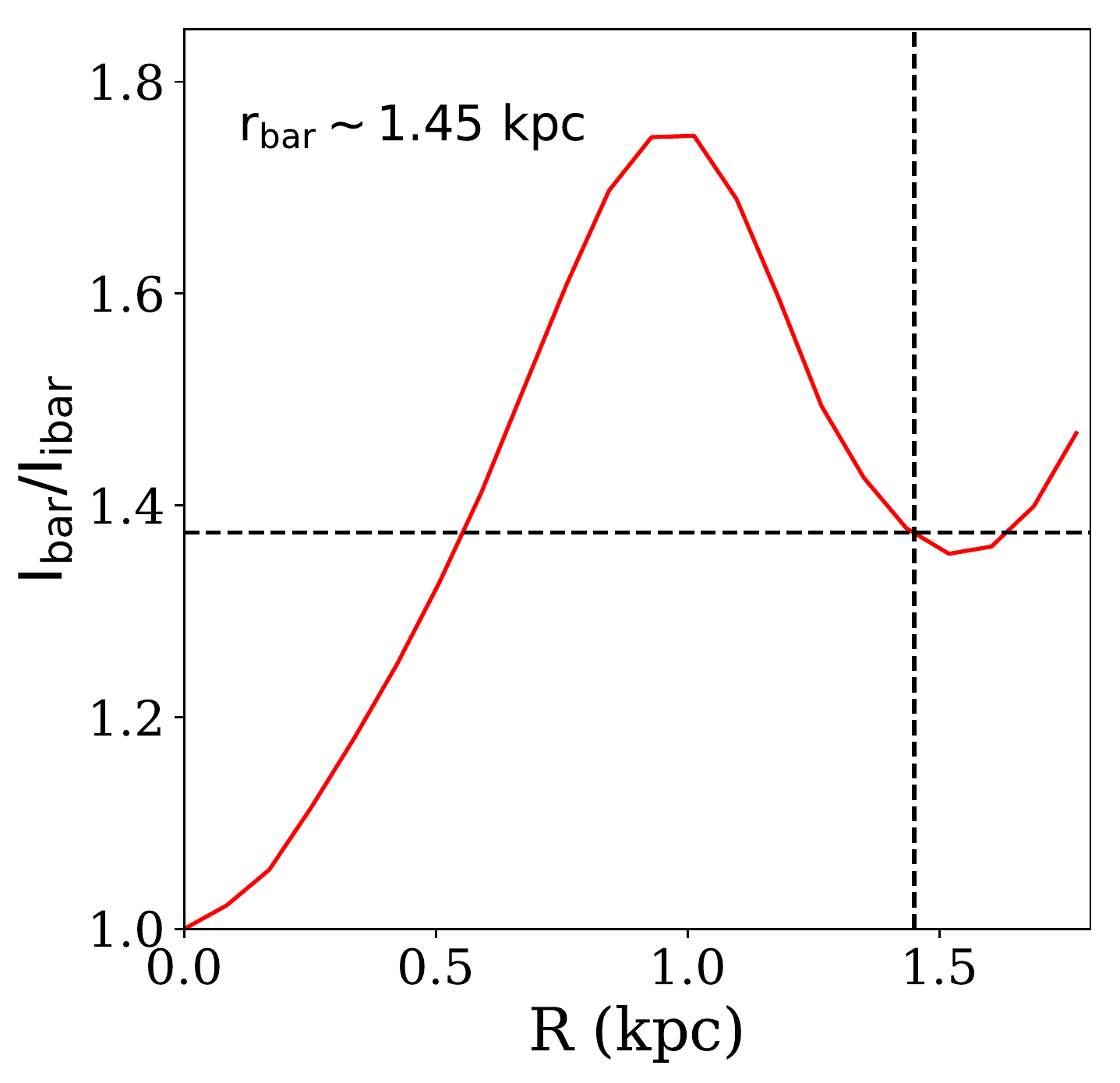}} 
\end{tabular}
\end{center}
\caption{Top panel: The de-projected Fourier amplitudes of different Fourier modes of the \HI~intensity distribution at the central region of DDO 168 as a function of radius. Bottom panel: Estimation of the \HI~bar length using Fourier decomposition method. The vertical axis represents the ratio of the bar to inter-bar intensity as a function of the deprojected radius in kpc. The black dashed horizontal line marks the condition of Eq.~\ref{eq_fourier}. The second intersect of Eq.~\ref{eq_fourier} with the $I_{bar}/I_{ibar}$ curve designates the bar radius in the galaxy plane as shown by the vertical dashed line. We find a bar radius, $R_b \sim $ 1.45 kpc using the Fourier decomposition method. See text for more details.} 
\label{fourier}
\end{figure}

\smallskip
\noindent{\it Fourier decomposition:} In this method, the deprojected azimuthal intensity profiles are decomposed into Fourier components or modes. These Fourier components are then further used to estimate the bar and the inter-bar intensity profiles and compared to locate the region where the bar intensity (m=2 mode) is higher than the inter-bar intensity. For example, if $I(r,\theta)$ denotes the de-projected azimuthal \HI~surface density profile, then it can be assumed to be the sum of all odd and even harmonics.

\begin{equation}
I(r,\theta) = \frac{A_0(r)}{2} \ +  \sum_{m=1,2,3,..} A_m(r) \cos (m\theta) \ + \ B_m(r) \sin (m\theta)
\end{equation}

\noindent where $A_0(r)$ represents the uniform intensity distribution across the galaxy (independent of $\theta$). The Fourier coefficients then can be given as

\begin{equation}
A_m(r) = \frac{1}{\pi} \int_0^{2\pi} I(r,\theta) \cos(m\theta) d\theta
\label{eq7}
\end{equation}
\begin{equation}
B_m(r) = \frac{1}{\pi} \int_0^{2\pi} I(r,\theta) \sin(m\theta) d\theta
\label{eq8}
\end{equation}

\noindent consequently, the Fourier amplitude of the $m^{th}$ component can be given as

\begin{equation}
I_m(r) = \sqrt{A_m^2(r) + B_m^2(r)}
\label{eq9}
\end{equation}

In Fig.~\ref{fourier} top panel, we plot the normalized Fourier amplitudes of different modes calculated using Eq.~\ref{eq7},~\ref{eq8} and~\ref{eq9}. At the central region where the bar intensity dominates, it is expected that the even modes will have significantly higher Fourier amplitude than the odd ones. However, from the figure, it can be seen that the $m=1$ Fourier mode is dominant at $R \lesssim 1.5$ kpc. In Fig.~\ref{visual} top panel, one can see, there exists a significant amount of lopsided emission on the side of the bar with positive $Y$ values. This could be the primary origin of enhanced m=1 or lopsided amplitude. While the fractional amplitude of m=1 seems to increase steeply with radius, in the region of the bar ($R \lesssim 1$ kpc), the m=1 and m=2 values are comparable, and in any case the m=1 component will not affect the even modes or the resulting bar parameters such as the bar length or strength. Note that when the m=1 mode has a high fractional amplitude (at $R > 1$ kpc) the actual strength of $I_1$ and $I_0$ are much smaller, so m=1 or lopsidedness is overall not a dynamically important component in the galaxy. In any case, the effect of lopsidedness stars to decrease beyond $R = 1.5$ kpc. The bar and inter-bar intensity can be defined as \citep[see, e.g.,][]{aguerri00}

\begin{equation}
I_{bar} = I_0 + I_2 + I_4 + I_6
\end{equation}
\noindent and,
\begin{equation}
I_{ibar} = I_0 - I_2 + I_4 - I_6
\end{equation}

The bar region then can be defined as where

\begin{equation}
\frac{I_{bar}}{I_{ibar}} > \frac{1}{2}\left((I_{bar}/I_{ibar})_{max} - (I_{bar}/I_{ibar})_{min}\right) + (I_{bar}/I_{ibar})_{min}
\label{eq_fourier}
\end{equation}

In Fig.~\ref{fourier} bottom panel we plot the $I_{bar}/I_{ibar}$ as a function of the deprojected radius in kpc. The horizontal black dashed line represents the RHS of Eq.~\ref{eq_fourier}. The vertical dashed line represents the outer radius where the condition of Eq.~\ref{eq_fourier} satisfies. Subsequently, we find the deprojected bar radius to be $\sim 1.45$ kpc. This value is somewhat higher than what is found using the other two methods. However, it is well-known that, if the bar structure is not tightly defined in its intensity distribution (which is the case for DDO 168), the Fourier decomposition method can significantly overestimate the bar length \citep[see, e.g.,][]{aguerri09,peters18,peters19}. For further calculations, we adopt a bar radius, $R_b = 1$ kpc. 

\subsection{Determination of bar strength}

The strength of the bar is one of the critical structural parameters used for the characterization of the bar. It represents the bar prominence and hence, the strength of the non-axisymmetric potential. The strength of a bar can be defined as \citep[][see also, \citet{peters19}]{ohta90}

\begin{equation}
s = \frac{1}{R_b} \sum_{m=2,4,6} \int^{R_b}_{0} \frac{I_m}{I_0} dr
\label{eq13}
\end{equation}

We use the amplitudes of the Fourier modes, as shown in the top panel of Fig.~\ref{fourier} to evaluate the above equation. We find the strength of the \HI~bar in DDO 168 to be $\sim 0.2$. This value of the bar strength is much lower than what is observed for strong bars in spiral galaxies \citep[see, e.g.,][and references therein]{guo19}. Hence, the \HI~bar in DDO 168 can be considered to be a weak bar. 

\subsection{\HI~bar is a slow bar}

Next, we estimate the ratio ($\mathcal{R}$) of the Lagrange point radius, $\rm R_{L}$ (or the corotation radius) to the bar radius, $\rm R_b$. The Lagrange radius or the co-rotation radius is the radius where the orbital speed of the particle in the galaxy disc is the same as the bar pattern speed. In other words at this radius, the rotational speed in the disc becomes zero when viewed from the reference frame of the bar. This is estimated by the maximum disc rotational velocity, $V_{max}$ (from the rotation curve) and the bar pattern speed, $\Omega_p$.

The ratio $\mathcal{R}$ is often used as an indicator to characterize the bar as a `fast' or a `slow' bar \citep[see, e.g.,][]{debattista00,sellwood06}. Dynamically, to maintain a stable orbit inside the bar, the ratio must be, $\mathcal{R} > 1$. For $1 \textless \mathcal{R} \textless 1.4$, the bar pattern speed is high resulting in a fast bar whereas, $\mathcal{R} > 1.4$ refers to a slow bar. However, recent studies indicate that, during the dynamical evolution of a galaxy, the bar also grows in length by accretion. In that case, the ratio $\mathcal{R}$, can acquire a value $<$ 1.4 and yet denote a slow bar \citep[see, e.g.,][]{font14a,font17}.

For DDO 168, $V_{max} = 61.9 \pm 0.8$ \kms~\citep{oh15}. Though for DDO 168, we find two different pattern speeds in two different halves, we use the mean pattern speed, i.e., $\Omega_p = 23.3 \pm 5.9$ \kmskpc~for the calculation of the corotation radius. With these values, we find $R_L = 2.7 \pm 0.1$ kpc and consequently an ${\mathcal R} = 2.7 \pm 0.7$. This value of ${\mathcal R}$ is much higher than the characteristic limit for a fast bar, and hence we conclude that the \HI~bar in DDO 168 is a slow bar. It should be emphasized that adopting a conservative approach, even if we consider the higher value of $\Omega_p = 30.0 \pm 7.2$ \kmskpc~in the lower half of Fig.~\ref{pspeed} (the rightmost panel), the ratio comes out to be, $\mathcal{R} = 2.1 \pm 0.5$ which is again an indicator of a slow bar. This, in turn, strongly indicates that the \HI~bar detected in DDO 168 is a slow bar for all conservative estimates.

Theoretically, it is expected that when the dark matter halo is dense \citep{debattista98} and hence dominates even in the inner parts, the dynamical friction between the bar and the density wake created by the rotating bar in the dark matter halo would significantly slow down the pattern speed resulting in a slow bar. It is sometimes said in the literature that LSBs or dwarf galaxies are dark-matter-dominated, this is a somewhat misleading statement since all galaxies are dominated by dark matter halo once one goes sufficiently far out in the galaxy. What is relevant is that the halo is dense and compact and hence the dark matter dominates even at inner radii as in LSBs \citep[e.g.,][]{ghosh14} or in dwarf galaxies like DDO 168 \citep{ghosh18}. It is this feature that influences the development of a slow bar due to dynamical friction. DDO 168 has a high central dark matter density, $\rho_0 = (39.81 \pm 6.37) \times 10^{-3}$ \mspcc and a scale radius, $r_s = 2.81 \pm 0.83$ kpc for an isothermal dark matter halo \citep{oh15}. It has been found that the dark matter halo dominates the observed dynamics of this galaxy starting from the central region \citep{ghosh18}. Thus the detection of a slow bar in DDO 168 supports the proposition that the dense, compact dark matter halo slows down the bar through dynamical friction.

We note that all the three galaxies that so far have been seen to host an \HI~bar (NGC 2915, NGC 3741 and DD0 168), also have the property that the bar is slow. We further note that the  \HI~gas is very extended (several times the optical radius) in all three galaxies, and in NGC 3741 and DDO 168 the halo is dense and compact \citep{ghosh18}.

Here it should be emphasized that now, there is evidence that slow bars are more common than what was thought before. For example, using an innovative and simple technique, \citet{font17} determined the corotation radius in 68 galaxies. Using the photometric data, they further determine the bar length in these galaxies and consecutively estimate the $\mathcal{R}$ to find that almost 1/3 of their sample galaxies host ``slow bars'' (i.e., $\mathcal{R} > 1.4$). This commonality of slow bars in galaxies can be thought of due to more effective braking by the dark matter halo than what was believed before. However, the status of the pattern speeds in dark matter dominated galaxies, especially LSBs, is not well studied as compared to the HSBs \citep{peters19,font17}. As shown in our works for DDO 168 and NGC 3741, the dense and compact halo might play an import role in slowing down the \HI~bars. Thus, though the braking of bars by dark matter halo is more common, yet the dark matter dominance from the central region might have an enhanced effect in determining the slowdown of the bar. 

Further, \citet{font14c} have shown by simulation that, due to the dynamical braking by the dark matter halo, the corotation radius moves outwards, at the same time, the bar can also grow in length significantly due to accretion.  This implies, even for a slow bar the ratio, $\mathcal{R} = R_C/R_b$ can settle well below the nominal value of 1.4, thus declaring the bar to be a fast one. Hence, all the fast bars as decided by $\mathcal{R} < 1.4$ criteria might not be fast ones. However, we clarify that the traditional definition of a slow bar (i.e., $\mathcal{R} \textgreater 1.4$) does not change our conclusion on the existence of the slow bar in DDO 168, albeit it puts questions on its rarity.

\section{Conclusion}

We have identified an \HI~bar in the dwarf galaxy DDO 168 which has a dense, compact dark matter halo which dominates from inner radii. This is only the third galaxy found to host an \HI~bar (the previous ones being NGC 2915 and NGC 3741). We apply the Tremaine-Weinberg method to the \HI~kinematic data from the LITTLE-THINGS survey and estimate the average pattern speed of the bar, $\Omega_p = 23.3 \pm 5.9$ \kmskpc. Interestingly, we also find different pattern speeds in the two different kinematic halves of DDO 168 which cannot be explained by the associated uncertainties in the measurements. Using simulated data and manipulating the kinematic axis we try to investigate the origin of the difference in the pattern speeds. We find that an error in the position angle and the asymmetry in the bar intensity distribution cannot produce the difference in the pattern speeds as observed. A kinematic asymmetry, i.e., an asymmetry in the velocity fields in the two kinematic halves is found to be the origin of such offset. We emphasize that this kind of observed difference in the pattern speeds puts a strong constraint on the lifetime of the observed \HI~bar. Employing basic dimensional analysis, we compute the lifetime of the \HI~bar in DDO 168 to be $\sim 5.3 \times 10^8$ years which is $\sim 2$ times the dynamical time scale ($\sim 2.6 \times 10^8$ years) of the disc. If \HI~bars being weak are easily disturbed then this could be a possible reason for the paucity of the \HI~bars in galaxies.

We further estimate the length of the \HI~bar in DDO 168 using three different methods, namely, the {\it visual inspection}, {\it position angle variation} and {\it Fourier decomposition}. Using the first two methods, we find a bar radius, $R_b \sim 1$ kpc, whereas the {\it Fourier decomposition} method yields a little higher bar radius of $\sim 1.45$ kpc. Using the values of the bar radius and the Lagrangian radius, we estimate the dimensionless ratio $\mathcal{R}=R_L/R_b$ to classify if the \HI~bar in DDO 168 is a `fast' or a `slow' bar. We find the value of $\mathcal{R}$ to be $\mathcal{R} \geq 2.1$ for all conservative determinations. This indicates that DDO 168 hosts a slow \HI~bar. Even though the existence of slow bars was proposed by \citet{debattista98} on theoretical grounds (as arising due to dynamical friction with the dark matter halo) over 20 years ago, detection of such bars in late-type galaxies is difficult and has only been done for three galaxies in literature so far (NGC 2915, NGC 3741 and F563-V2). Thus detection of a slow bar in a late-type galaxy as done here is interesting in itself, and thus DDO 168 is an important addition to a small set of late-type galaxies showing slow bars. However, we note that \citet{font17} using another method to determine the corotation radius find that nearly 30\% of 68 HSB galaxies they study have $\rm \mathcal{R} \textgreater 1.4$, thus have slow bars. This confirms that slow bars may not be as rare as once believed.


\section{Acknowledgement}
We thank the referee, John Beckman, for his comments and suggestions that have greatly helped to improve the quality of the paper. NNP would like to thank Deidre Hunter and Fabian Walter for sharing the visibility data of DDO 168 which we used extensively for this work. CJ would like to thank the DST, Government of India for support via a J.C. Bose fellowship (SB/S2/JCB-31/2014).

\bibliographystyle{mn2e}
\bibliography{bibliography}

\end{document}